\documentclass[%
 reprint,
 amsmath,amssymb,
 aps,
]{revtex4-1}

\usepackage{graphicx}
\usepackage{psfrag}
\usepackage{dcolumn}
\usepackage{bm}

\usepackage{hyperref} 
\usepackage{xcolor}
\hypersetup{
    colorlinks,
    linkcolor={red!50!black},
    citecolor={blue!70!black},
    urlcolor={blue!90!black}
}
\begin{document}

\preprint{APS/123-QED}

\title{de Gennes Narrowing and the Relationship Between Structure and Dynamics in Self-Organized Ion Beam Nanopatterning}

\author{Peco Myint}
 \email{peco@bu.edu}
\affiliation{%
Division of Materials Science and Engineering, \\Boston University, Boston, Massachusetts 02215 USA}

\author{Karl F. Ludwig, Jr.}%
 \email{ludwig@bu.edu}
\affiliation{%
Department of Physics and\\Division of Materials Science and Engineering, \\Boston University, Boston, Massachusetts 02215 USA}
\author{Lutz Wiegart}
\author{Yugang Zhang}
\author{Andrei Fluerasu}
\affiliation{
 National Synchrotron Light Source II, \\Brookhaven National Lab, Upton, NY 11973 USA
}%

\author{Xiaozhi Zhang}
\author{Randall L. Headrick}
\affiliation{
 Department of Physics and Materials Science Program, \\University of Vermont, Burlington, Vermont 05405 USA
}%

\date{\today}

\begin{abstract}
{\noindent 

Investigating the relationship between structure and dynamical processes is a central goal in condensed matter physics. Perhaps the most noted relationship between the two is the phenomenon of de Gennes narrowing, in which relaxation times in liquids are proportional to the scattering structure factor.  Here a similar relationship is discovered during the self-organized ion beam nanopatterning of silicon using coherent x-ray scattering.  However, in contrast to the exponential relaxation of fluctuations in classic de Gennes narrowing, the dynamic surface exhibits a wide range of behaviors as a function of length scale, with a compressed exponential relaxation at lengths corresponding to the dominant structural motif - self-organized nanoscale ripples.  These behaviors are reproduced in simulations of a nonlinear model describing the surface evolution.  We suggest that the compressed exponential behavior observed here is due to the morphological persistence of the self-organized surface ripple patterns which form and evolve during ion beam nanopatterning.
}

\end{abstract}

\maketitle


Understanding the relationship between structure and dynamical processes is a central goal in condensed matter physics.  Coherent x-ray scattering has emerged as a powerful approach which can simultaneously investigate dynamics through X-ray Photon Correlation Spectroscopy (XPCS) and structure through measurement of the scattering structure factor \cite{sutton2008review,madsen2015structural}.  XPCS is a speckle correlation technique analogous to dynamic light scattering (DLS) \cite{berne2000dynamic}, but using x-rays to reach smaller length scales than is possible with visible light.  The quintessential relationship between fluctuation dynamics and structure is the phenomenon of de Gennes narrowing, in which the neutron inelastic scattering linewidth $w(q)$ in liquids is found to narrow at the position of the peak in the scattering structure factor $I(q)$, with $w(q) \sim I(q)^{-1}$ \cite{degennes1959}.  The corresponding correlation times $\tau(q)$ are then proportional to $I(q)$. Here we use coherent x-ray scattering to explore the relation between structure and dynamics during the widespread phenomenon of self-organized ion beam nanopatterning, during which partly ordered surface structures spontaneously appear during broad beam ion bombardment \cite{baglin2020ion}.  Telling similarities with de Gennes narrowing are observed, but also important differences.  Simulations of the surface evolution are shown to reproduce the observed behavior which we relate to the evolving ripple morphology on the surface.

Speckle correlation experiments such as XPCS measure dynamics in the time domain through correlation functions, particularly the intensity autocorrelation $g_2(\textbf{q},\Delta t) = \left< I(\textbf{q},t_1)I(\textbf{q},t_2) \right>_{t'} / \left< I(\textbf{q},t') \right>_{t'} ^2$, where $\Delta t = |t_1 - t_2|$ and $t'$ represents the range in $t_1$ and $t_2$ times over which the function is calculated\cite{berne2000dynamic}.  For a practical matter, this function is also typically averaged over equivalent values of $\textbf{q}$.  While the detailed behavior of the correlation function can be complex, experimentally they can often be fit with a Kohlrausch-Williams-Watts (KWW) function \cite{williams1970non}: $g_2(\textbf{q},\Delta t) = \{ 1 + \beta \textrm{exp} [-2(\Delta t/\tau(\textbf{q}))^{n(\textbf{q})}] \}$, where $n(\textbf{q})$ is an exponent characterizing the fluctuation decay and $\beta$ is a setup dependent contrast factor, determined by the coherence properties of the x-ray beam and the spatial sampling of the speckles \cite{sandy2018hard}. 

It's conventional to describe structure and dynamics in terms of a spatially-dependent system ordering variable $\phi(\textbf{r},t)$.  For a liquid $\phi(\textbf{r},t)$ represents atomic density fluctuations $\rho(\textbf{r},t)$ while for ion beam nanopatterning it usually represents surface height fluctuations $h(\textbf{r},t)$.    In a simple linear process with uncorrelated Gaussian random noise $\eta(\textbf{r},t)$, the evolution of order in reciprocal space $\phi(\textbf{q},t)$ is:
\begin{equation}
\frac{\partial \phi (\mathbf{q},t)}{\partial t} = - A(\mathbf{q}) \phi(\mathbf{q},t) + \eta(\mathbf{q},t).
\label{equ:lineq}
\end{equation}
If $A(\textbf{q})$ is positive so that the system is stable to fluctuations, then, in equilibrium, the structure factor intensity is $\langle I(\textbf{q}) \rangle \sim \langle \phi^*(\textbf{q},t)\phi(\textbf{q},t) \rangle \sim \langle \eta^2 \rangle / \it{A}(\textbf{q})$. The intensity autocorrelation function $g_2(\textbf{q},\Delta t)$ exhibits an exponential decay with time constant 1/2$A(\textbf{q})$ \cite{bagchi2001mode} so that $\tau(\textbf{q}) \sim \langle \it{I}(\textbf{q}) \rangle$. Thus, the classic de Gennes relationship is a hallmark of a purely linear dynamics.  If the dynamics contains nonlinearities then the relaxation is no longer exponential and the deviation of the KWW exponent $n(q)$ provides an important characterization of the nonlinear behavior.
\begin{figure}
\includegraphics[width=3.2 in]{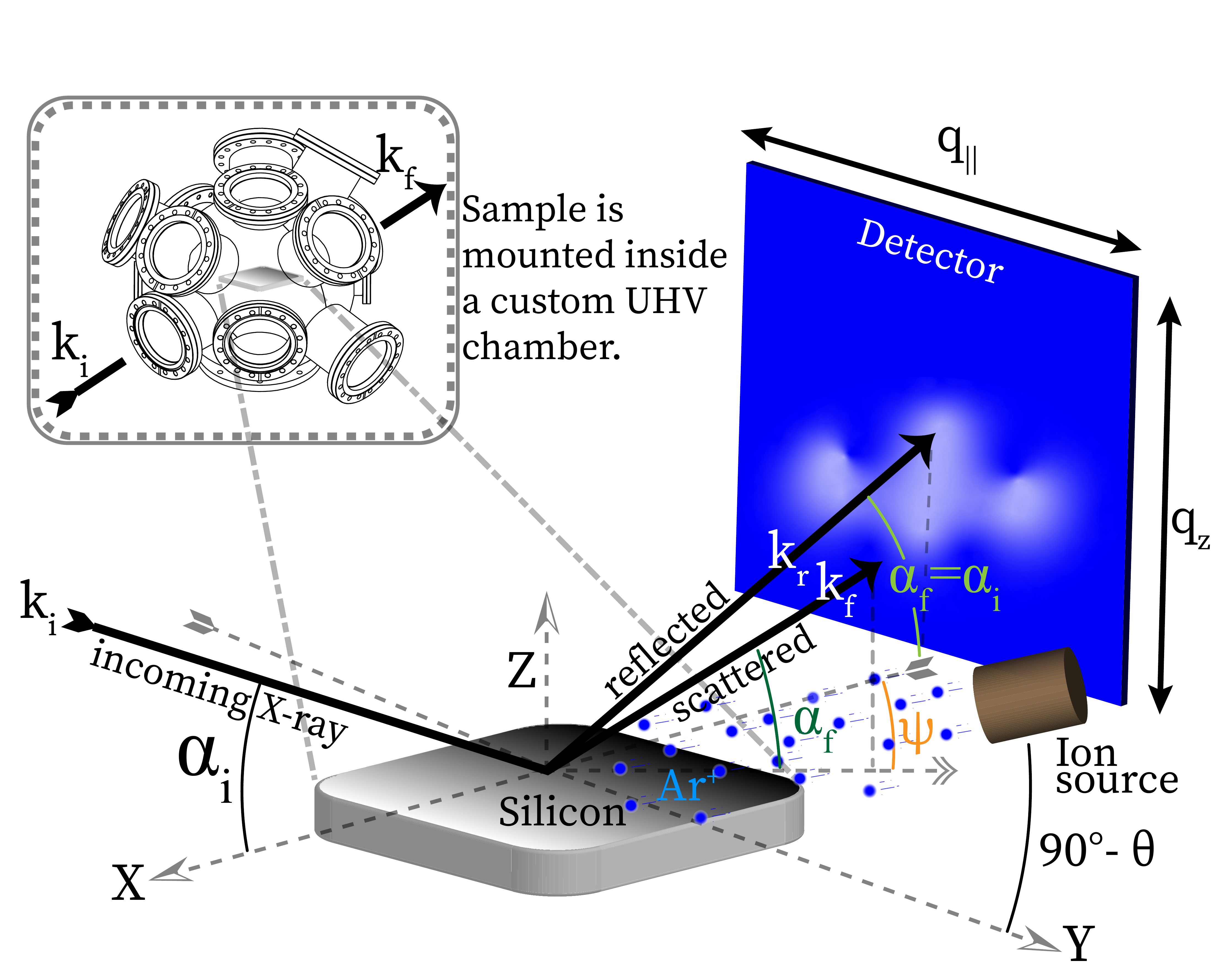}

\includegraphics[width=3.2 in]{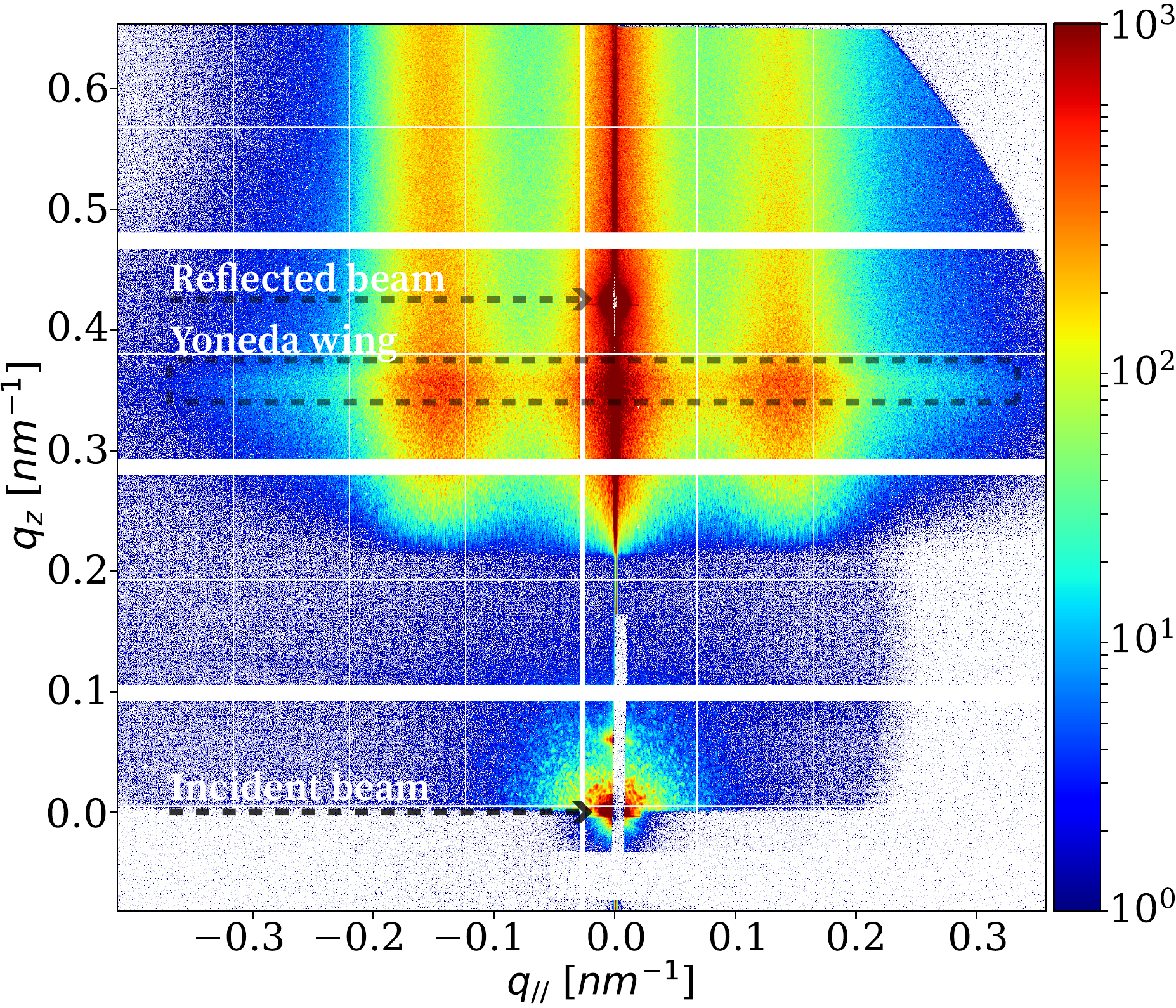}
\caption{\label{fig:GISAXS} Top: A schematic diagram of the GISAXS experiment. The ion source is placed at the polar angle $\theta$, which causes self-organized rippling on the silicon surface. The sample is positioned so that the X-ray incident angle $\alpha_i$ is slightly above the critical angle of total external reflection. The scattering is recorded as a function of the exit angles $\alpha_f$  and $\psi$ using a 2D detector. Bottom:  A detector image during nanopatterning. The Yoneda wing, spread across $q_{z}\,=\,0.36\,nm^{-1}$ or $q_{z}^\prime\,=\,0.16\,nm^{-1}$, is the surface-sensitive scattering exiting the sample at the critical angle, $\alpha_c$. For Ar$^+$ patterning, correlation peaks at $q_{||}\,=\, q_0\simeq\pm\,0.18\,nm^{-1}$ are due to the correlated nanoripples on the surface.}
\end{figure}
 
In order to study the nanoscale evolution of surfaces during ion beam patterning, this study uses a  surface-sensitive Grazing Incidence Small-Angle X-ray Scattering (GISAXS) geometry.  The classic cases of 1 keV Ar$^+$ and Kr$^+$ ion bombardment of silicon are studied at an ion incidence angle of $\theta = 65^{\circ}$. The two ions cause generally similar patterning behavior, and we find in this study that their dynamics behavior is similar, suggesting that the phenomena reported here may be widespread. At the chosen ion bombardment angle, nano-ripples are known to form on the surface with a wavevector parallel to the projected direction of the incoming ions onto the sample surface \cite{norris2017distinguishing,perkinson2018sawtooth}.  Experiments used a photon energy of 9.65 keV and the incident x-ray angle $\alpha_i$ was 0.26$^{\circ}$, which is slightly above the critical angle of total external reflection.  Detector images were recorded every 0.25 s. In the x-ray scattering, the direction parallel to the ripple wavevector is denoted $q_{||}$ and the direction perpendicular to the surface is $q_z$.  These correspond approximately with the horizontal and vertical directions on the position-sensitive x-ray detector (Fig. \ref{fig:GISAXS}). Ions enter from the side of the positive $y$-axis. 

We focus primarily on x-ray scattering intensity along the Yoneda wing, which is particularly sensitive to surface morphology \cite{santoro2017grazing}.  In the limit that $q_z h$ is small, the GISAXS intensity along the Yoneda wing is proportional to the height-height structure factor \cite{sinha1988x}. As $h$ becomes larger, the scattering intensity is closely related to the height-height structure factor, but reduced by destructive interference within structures.  As shown in the bottom part of Fig. \ref{fig:GISAXS}, during bombardment correlation peaks grow in the scattering pattern at wavenumbers $ \pm q_0$; these are due to the formation of the correlated nano-ripples with wavelength $\lambda = 2 \pi / q_0$.  The ripple structure coarsens with time, but at a very slow rate so that the correlation peaks remain well within the detector window throughout the experiment.  For our purposes here, this is an important difference with previous XPCS studies of ion beam nanopatterning of GaSb \cite{bikondoa2013ageing} and SiO$_2$ \cite{mokhtarzadeh2019nanoscale}.  It's known that the ripples move across the surface during continued bombardment and that, at late times, the sinusoidal ripples develop into asymmetric sawtooth structures \cite{gago2002nanopatterning,engler2014evolution,basu2013transition,pearson2014theory}, which creates asymmetry in the scattering between positive and negative $q_{||}$ \cite{ludwig2002si,perkinson2018sawtooth}.

 \begin{figure}
\includegraphics[width=3.4 in]{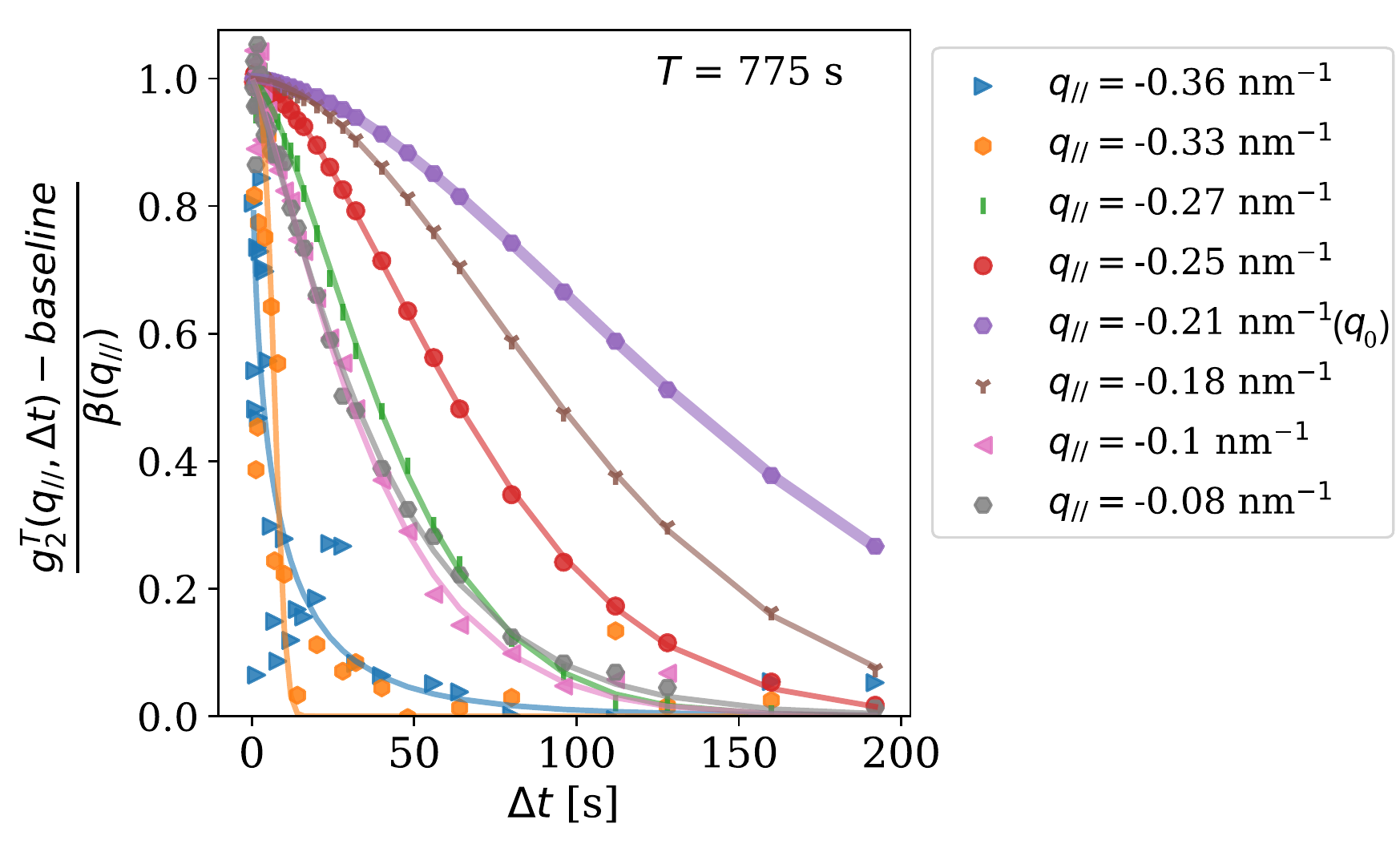}%
\caption{\label{fig:Kr_g2_linearT775}KWW fits of $g_2^T(q_{||},\Delta t)$ for select wavenumbers at aging time $T$=775 s. For Kr$^+$ patterning, correlation peaks are at $q_{||}\,=\, q_0\simeq\pm\,0.21\,nm^{-1}$.}
\end{figure}
 To quantitatively analyze the evolution, age-dependent correlation functions $g_2^T(q_{||},\Delta t)$ are calculated with the range t' over which the autocorrelation function is evaluated being $\approx \pm 125$ s around a central aging time $T = (t_1 + t_2)/2$ \cite{bikondoa2017use}.  The value of $\approx \pm 125$ s was chosen to give sufficient $\Delta t$ over which to evaluate the decay of correlations while still being relatively small compared to the overall evolution of the sample. The age-dependent correlation functions were fit with the KWW form to determine age-dependent correlation times $\tau(q_{||})$ and exponents $n(q_{||})$.  Figure \ref{fig:Kr_g2_linearT775} shows examples of the normalized $g_2^T(q_{||},\Delta t)$ data and fits at various negative wavenumbers for $T = 775 \: \mathrm{s}$. A clear change from compressed exponential behavior ($n > 1$) near the ripple wavenumber $q_0 \simeq -0.21 \mathrm{nm}^{-1}$ (as evidenced by downward curvature at $\Delta t=0$) to stretched exponential behavior ($n < 1$ at high wavenumbers (as evidenced by upward curvature at $\Delta t=0$) is observed. 
 
\begin{figure}
\includegraphics[width=3.4 in]{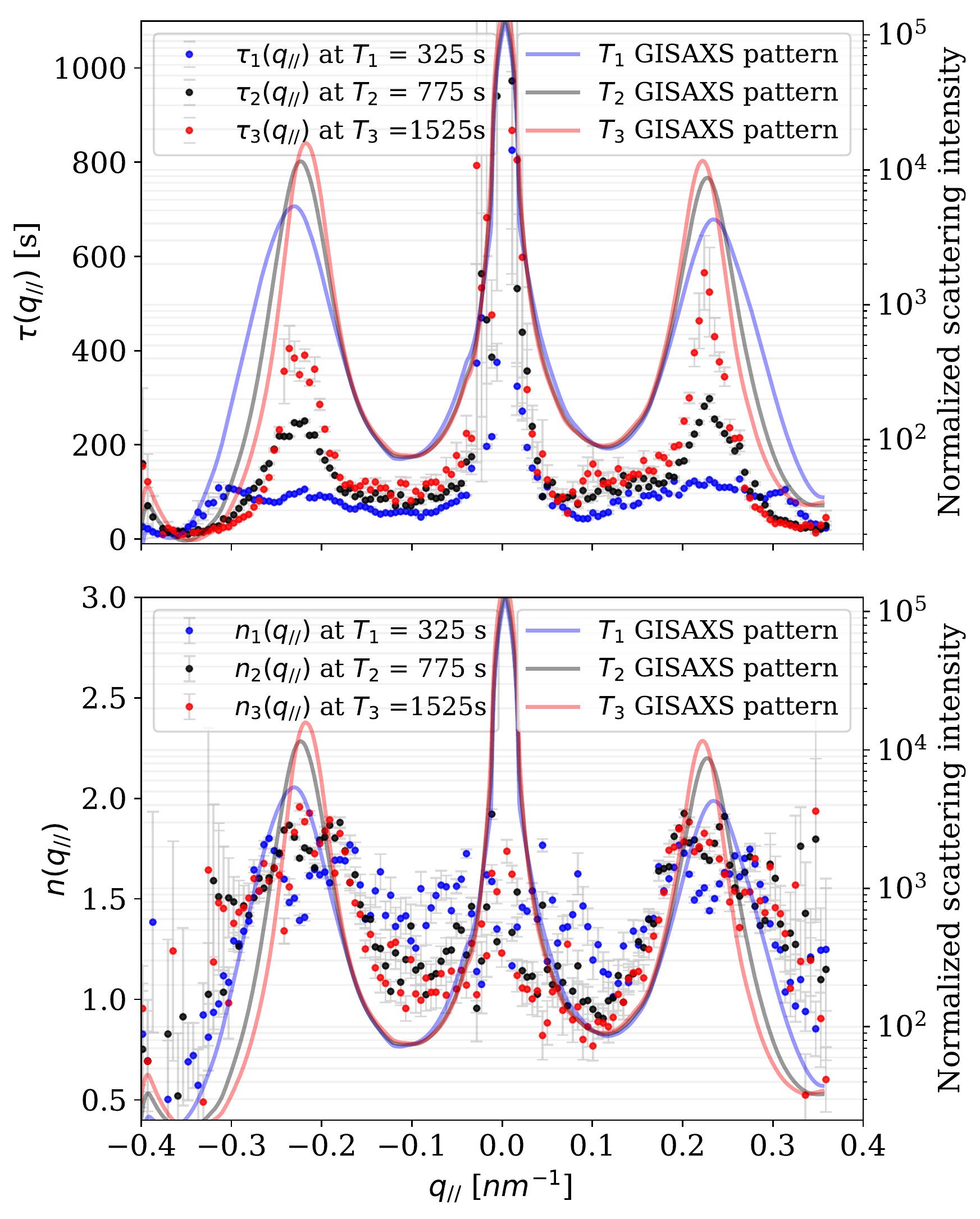}%
\caption{\label{fig:Kr_3g2_exponents} Correlation times $\tau(q_{||})$ and exponents $n(q_{||})$ extracted from $g_2^T$ fits at different aging times $T$ for Kr$^+$ patterning. Normalized $I(q_{||})$ plots at corresponding aging times $T$ are also shown.}
\end{figure}

 Figure \ref{fig:Kr_3g2_exponents} exhibits how $\tau(q_{||})$ and $n(q_{||})$ evolve with $T$ for Kr$^+$ bombardment, in comparison with the speckle-averaged scattering itself which shows peaks due to the continued growth and coarsening of correlated ripple structures. At the earliest times shown, there is only a modest variation of $\tau(q_{||})$ with wavenumber for $|q_{||}| < 0.32 \; \mathrm{nm}^{-1}$.  With continued evolution, $\tau(q_{||})$ grows markedly at the ripple wavenumber, forming an increasingly narrow peak.  The speckle-averaged intensity $I(q_{||})$ is slightly higher at negative $q_{||}$ than at positive $q_{||}$, while $\tau(q_{||})$ behaves in the opposite manner. Figure \ref{fig:Kr_3g2_exponents}  also shows that the KWW exponent $n(q_{||})$ exhibits rich behavior, developing a peak near the ripple wavenumber with a maximum value of approximately 1.7-1.8, but falling to a value well below one at high wavenumbers. 

To better understand the observed behavior, we have performed simulations of a nonlinear model describing ion beam nanopatterning due to Harrison, Pearson and Bradley (HPB) \cite{pearson2014theory,harrison2017emergence}:
\begin{eqnarray}
\frac{\partial h(\textbf{r},t)}{\partial t} = &&A \, h_y +\nu _x \, h_{xx} + \nu _y \, h_{yy} + \lambda _x \, h_x^2 +\lambda _y \, h_y^2 + \nonumber\\&&\gamma _y \,h_y^3 -\kappa\nabla^{4}h +\eta(\textbf{r},t).
\label{equ:HPB}
\end{eqnarray}
where $\eta(\textbf{r},t)$ is a Gaussian white noise and the individual coefficients and terms are related to detailed surface response. Within the HPB equation, the first derivative term is related to the angle-dependence of the sputter erosion and its primary effect is simply to drive structures across the surface. The curvature terms are due to the curvature-dependent sputter yield and lateral mass redistribution. In our case, $\nu_y < 0$, $\nu_x > 0$, so that the initial surface is unstable to the formation of ripples in the $y$-direction, which is the direction probed by the x-ray experiments. The $\nabla^{4}h$ term is a long-wavelength approximation for surface viscous flow.  The nonlinear terms are higher order effects associated with the angle-dependence of the sputter yield.  The effect of the quadratic terms is to control the exponential growth of the instability, while the $h_y^3$ term causes the initial sine-wave ripple structure to evolve into an asymmetric sawtooth structure and to coarsen.  Numerical integrations were performed on a 2048 $\times$ 2048 lattice and, for comparison with experiment, the lattice size and time units in the simulations are set as 1 nm and 1 s respectively.  Coefficients used in the simulations were chosen to reasonably model the Ar$^+$ bombardment: $A$ = -0.26, $S_x$ = 0.45, $S_y$ = -0.45, $B$ = 6.96, $\lambda_x$ = 1.94, $\lambda_y$ = 1.94, $\gamma_y$ = 11.89, $<\eta^2>$ = 0.1. 

\begin{figure}
\includegraphics[width=3.4 in]{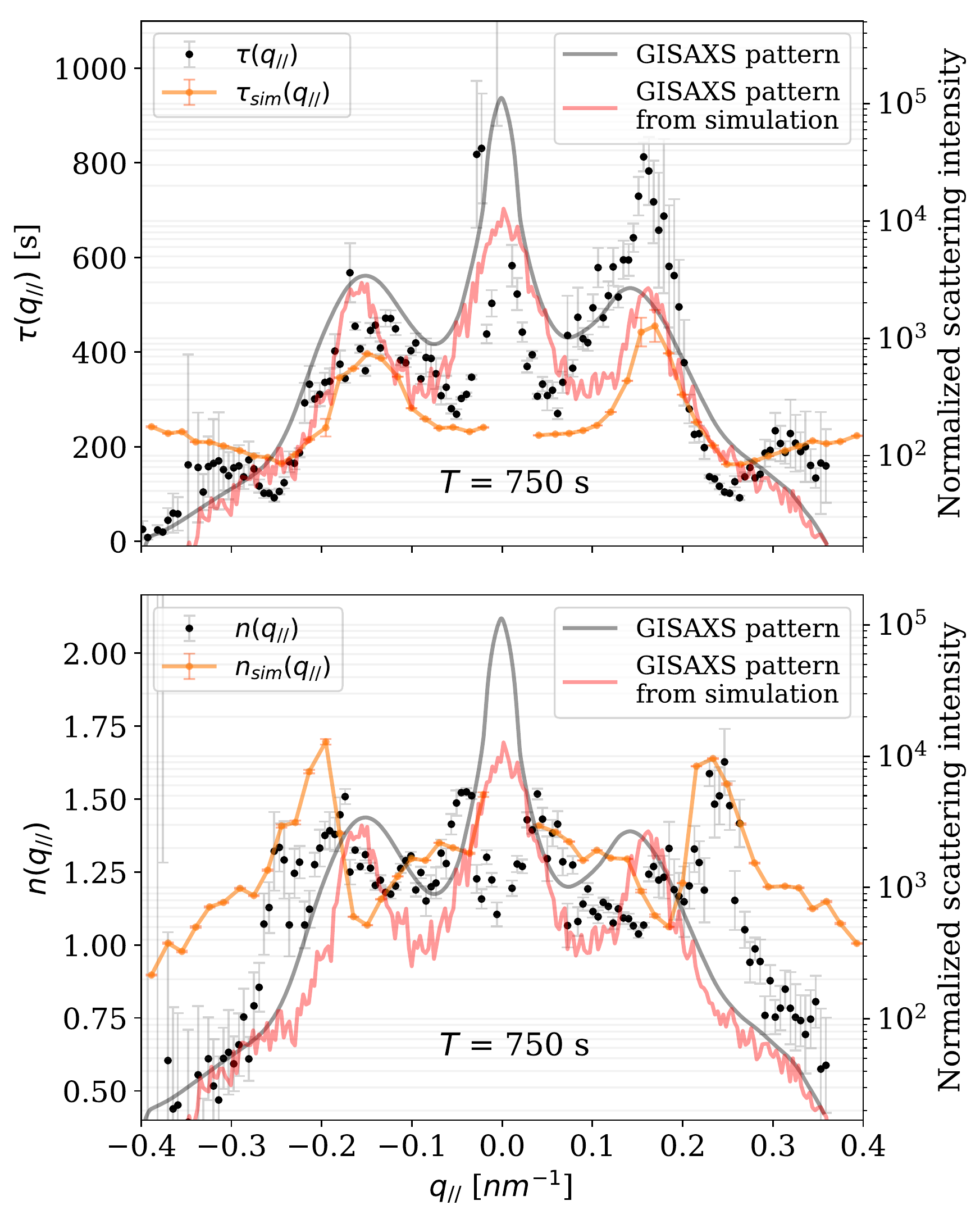}%
\caption{\label{fig:Ar_exponents_and_simulations} Correlation times $\tau(q_{||})$ and exponents $n(q_{||})$ extracted from $g_2^T$ fits in Ar$^+$ patterning and its simulation.}
\end{figure}

Predicted GISAXS scattered intensities $I_{sim}(q_{||})$ were calculated from the simulations using the formalism of Sinha \textit{et al.} \cite{sinha1988x}.  From these, intensity autocorrelation functions $g_{2-sim}^T(q_{||},t)$ were calculated and fit to determine $\tau_{sim}(q_{||})$ and $n_{sim}(q_{||})$ at $T = 750 \textrm{s}$.  These are shown in Fig. \ref{fig:Ar_exponents_and_simulations}.  A good overall agreement with the experimental trends is observed.  In particular, an asymmetry in $I(q)$ develops, with the peak being slightly higher on the left than on the right side.  This effect can also be calculated directly from asymmetric sawtooth patterns when the slope is higher on the positive side (ion beam) side of the sawtooth than on the negative side.  This is indeed what is observed in analysis of \textit{post-facto} AFM images and is consistent with literature results. In addition, $\tau_{sim}(q_{||})$ shows peaks at the ripple wavenumbers $\pm q_0$, as does the exponent $n(q_{||})$.

The correspondence between peaks in the scattering intensity and in $\tau(q_{||})$ found during nanopatterning is evocative of the relation between structure and dynamics embodied in de Gennes narrowing, despite the great difference in their physical settings. While the correlation time and scattering intensity are simply proportional in classic de Gennes narrowing, in the present case the relative variation in $\tau(q_{||})$ is much less than the relative variation in $I(q_{||})$ as Fig. \ref{fig:Kr_3g2_exponents} shows.  Concomitantly, $g_2^T(q_{||},\Delta t)$ near the ripple wavenumber exhibits strongly compressed exponential behavior.

It's worthwhile to examine these behaviors more closely.  In liquids, de Gennes narrowing is associated with time scales of alpha-relaxation local diffusive motion; these time scales can range from picoseconds in atomic liquids to seconds and beyond in complex fluids and glasses.  In contrast, the time scale measured in the surface nanopatterning is that required for processes associated with the ion bombardment to change the surface morphology on a given length scale - particularly the length scale of surface nanostructures which are tens of nanometers in this case.

Turning to the observed non-exponential dynamics, for times $\Delta t < \tau$ a compressed exponential is a \textit{slower} decay than a simple or stretched exponential having the same time constant.  The limiting slope of the KWW function as $\Delta t \rightarrow 0$ is proportional to $(\Delta t)^{n-1}$, so that stretched exponential behavior at early times requires a large contribution from short decay times, while a compressed exponential behavior $n > 1$ at early times suggests their absence \cite{bouchaud2008anomalous}.  The most common rationale cited for structural persistence causing compressed exponential behavior at short times $\Delta t$ in both soft materials \cite{cipelletti2005slow} and metallic glasses \cite{ruta2012atomic} is collective ballistic flow of local structures due to internal stress relaxation. While the $\nabla ^4 h$ term in the HPB equation can model surface viscous relaxation, it’s a linear term which does not itself lead to nonexponential dynamics.  It's noteworthy that alternative theoretical approaches to understanding ion beam nanopatterning exist using fluid dynamic models with stress relaxation as a driving force \cite{castro2012hydrodynamic,norris2012stress,moreno2015stress,munoz2019stress}. These might provide a direct connection between the compressed exponential behavior of ion beam nanopatterning observed here and that observed in glasses.  This issue deserves further investigation.

These points notwithstanding, compressed exponential dynamics is not limited to systems in which stress-driven relaxation plays a critical role. In phase ordering and phase separating systems \cite{brown1997speckle, brown1999evolution, livet2001kinetic, fluerasu2005x, ludwig2005x} the theoretical form for the decay of correlations involves modified Bessel functions which numerical calculation shows can be well fit with the KWW function for $\Delta t < 2 \tau$, giving $n \approx 1.56$ and $n \approx 1.69$ for 2- and 3-dimensions respectively.  The compressed exponent is linked to the persistence of the domain structure \cite{brown1997speckle,brown1999evolution}. Temporal and spatial persistence has also been extensively studied for the Kardar-Parisi-Zhang (KPZ) growth model \cite{kardar1986dynamic} using various metrics and the HPB equation does include a KPZ-like quadratic nonlinearity $h_x^2,h_y^2$.   For small $\Delta t$, simulations of the KPZ model are well fit with compressed exponential behavior \cite{mokhtarzadeh2017simulations} and the leading terms in the KPZ model dynamics \cite{katzav2004numerical} suggest an effective exponent $n \approx (2+2\alpha)/z \approx 1.74$, where $\alpha$ and $z$ are the roughness and dynamic exponents respectively for (2+1) dimensional growth. Thus the KPZ effective compressed exponent $n$ for small $\Delta t$ is approximately equal to that observed at the nanoripple wavenumber peaks $\pm q_0$ in these experiments.  In the HPB equation, the $h_{xx}$ and $\nabla^4h$ terms are necessary to produce ordering, as opposed to only the kinetic roughening of the pure KPZ equation.  However, given the similarities in dynamics behavior between the HPB and KPZ equations, it seems likely that the $h_x^2$ nonlinearity plays a major role in determining the dynamics.  It's perhaps also notable that in (2+1) dimensions, the KPZ model has a predicted intensity variation in wavenumber $I(q) \sim q^{-2-2\alpha} \approx q^{-2.8}$ which is significantly larger than the variation in time constant $\tau(q) \sim q^{-z} \approx q^{-1.6}$, just as the situation found here for the nanopatterning.

In qualitative analogy with the persistence exhibited in phase ordering/separation and KPZ systems, here we conjecture that the compressed exponential dynamics observed near $\pm q_0$ points to the local morphological stability of the ripple pattern even as it moves across the surface.  At higher wavenumbers $|q| > q_0$, the KWW exponent decreases to one and below, suggesting that this persistence does not exist on shorter length scales; instead there is a range of relaxation rates, including some which are quite short.  To better quantify these issues it would be interesting to expand theoretical inquiry into the persistence of surface morphology using metrics that are applicable to the coherent scattering investigations of dynamics now possible.

This material is based on work supported by the National Science Foundation (NSF) under Grant No. DMR-1709380. X.Z. and R.H. were supported by the U.S. Department of Energy (DOE) Office of Science under Grant No. DE-SC0017802. Experiments were performed at the Coherent Hard X-ray (CHX) beamline at National Synchrotron Light Source II (NSLS-II), a U.S. Department of Energy (DOE) Office of Science User Facility operated for the DOE Office of Science by Brookhaven National Laboratory under Contract No. DE-SC0012704. We thank Glenn Thayer and Heitor Mourato of the Boston University Scientific Instrumentation Facility for design and construction of the sample holder.  We also thank comments from our anonymous reviewers, which made a substantive impact on the discussion and conclusions.

\bibliography{ionpatterning_references.bib}

\end{document}